\documentclass[acmsmall,screen,nonacm]{acmart}

\usepackage[T1]{fontenc}
\usepackage[utf8]{inputenc}
\usepackage{amsmath,amsfonts}
\usepackage[linesnumbered,ruled,vlined]{algorithm2e}
\usepackage{textcomp}
\usepackage{subfigure}
\usepackage{enumitem}
\usepackage{xspace}
\usepackage{pifont}
\usepackage{acronym}
\usepackage{graphicx}
\usepackage[nobiblatex]{xurl}
\usepackage{tabularx}

\usepackage[most]{tcolorbox}

\newtcolorbox{promptbox}[1]{
    enhanced,
    colback=gray!4,
    colframe=gray!45,
    coltitle=white,
    colbacktitle=gray!50,
    title=\textbf{#1},
    fonttitle=\bfseries,
    boxrule=0.8pt,
    arc=1.5mm,
    left=2.5mm,
    right=2.5mm,
    top=1.5mm,
    bottom=1.5mm,
    before skip=0pt,
    after skip=0pt
}

\renewcommand{\checkmark}{\ding{51}}
\newcommand{\xmark}{\ding{55}}

\acrodef{gas}[GAS]{Graph-as-State}

\begin{document}

\title{STAIR: Effective Incident Response Using an End-to-End Agentic Planning Framework}

\author{Hanlin Jiang}
\affiliation{%
  \institution{Peking University}
  \country{China}
}

\author{Jionghao Huang}
\affiliation{%
  \institution{Southeast University}
  \country{China}
}

\author{Shaofei Li}
\affiliation{%
  \institution{Peking University}
  \country{China}
}

\author{Bojia Yu}
\affiliation{%
  \institution{Southeast University}
  \country{China}
}

\author{Peng Jiang}
\affiliation{%
  \institution{Southeast University}
  \country{China}
}

\author{Yuxin Ren}
\affiliation{%
  \institution{Huawei Technologies Co., Ltd.}
  \country{China}
}

\author{Ning Jia}
\affiliation{%
  \institution{Huawei Technologies Co., Ltd.}
  \country{China}
}

\author{Yao Guo}
\affiliation{%
  \institution{Peking University}
  \country{China}
}

\author{Ding Li}
\affiliation{%
  \institution{Peking University}
  \country{China}
}

\begin{abstract}
Incident response planning is critical for restoring compromised software systems.
Existing AIOps methods automate fault diagnosis and remediation and adapt response using feedback, but lack a unified decision basis across planning and execution.
This limits their ability to maintain incident state and align actions with evolving recovery objectives during long-horizon, multi-stage security incident response.
We present STAIR, an end-to-end agentic response framework that organizes planning and execution around a shared incident state.
STAIR maintains incident evidence, response actions, and execution feedback in Graph-as-State (GAS), using the latest state to determine the response stage and recovery objectives for stage-specialized planning.
The framework retrieves effect-validated historical experience based on the current state and stage to guide action selection.
An Execution Harness executes actions, returns feedback to update GAS, and validates action effects for future experience reuse.
We construct STAIR-Bench, an executable incident response benchmark comprising 44 Docker-based cyber-range cases.
On its held-out test cases, STAIR achieves an overall defense score of 0.95, improving over the strongest baseline by approximately 10.47\%.
In the independent CAGE4 environment with ongoing attacks and benign activity, STAIR reduces cumulative penalties by approximately 28.8\% relative to the strongest baseline.
\end{abstract}

\maketitle

\section{Introduction}
\label{sec:introduction}

Security incident response is critical to maintaining the continuity of modern software services~\cite{kokulu2019matched,jones2023work}.
Detection identifies a potential compromise, but limiting its impact depends on timely investigation, containment, removal of attacker footholds, and service restoration.
IBM reports that the global average cost of a data breach fell to \$4.44 million in 2025, attributing part of the decline to faster identification and containment~\cite{ibm2025breach}.
The 2024 Change Healthcare cyberattack further illustrates the operational consequences: UnitedHealth reported \$2.2 billion in direct response costs and an estimated \$867 million in business-disruption impacts for the year~\cite{unitedhealth2024annual}.

Automating security incident response is difficult because recovery must be coordinated with the suppression of adversarial capabilities that may persist across multiple response stages.
Unlike conventional faults or service outages, restoring availability does not imply that a security incident has been resolved: stolen credentials, persistence mechanisms, malicious artifacts, or exploitable attack paths may remain~\cite{nelson2025incident,wu2026cutting}.
Meanwhile, actions such as host isolation, process termination, artifact removal, and service restart may suppress attacker capabilities while disrupting legitimate services or destroying forensic evidence~\cite{shaked2023operations,schlette2024you}.
Response objectives also change across containment, assessment, preservation, eviction, hardening, and restoration.
Automated response therefore requires iterative observation, intervention, and effect verification over a running software system~\cite{yao2022react,weyns2023activforms,calinescu2010dynamic}.

Recent AIOps research has expanded from fault diagnosis and service remediation toward automated handling of complex runtime incidents~\cite{sarda2024leveraging,fu2025llm,lin2026actionnexvirtualoutagemanager,chen2026stratus}.
E2E-REME~\cite{zhang2026e2e} connects microservice diagnosis with executable remediation, Triangle~\cite{yu2025triangle} supports production incident triage through a multi-agent workflow, and GALA~\cite{tian2026gala} uses service-dependency graphs to organize runtime evidence and generate response recommendations.
This trend also extends to security incident response: Hammar et al.~\cite{hammar2025incident} generate response actions from incident context and recovery state, SOAR4BC~\cite{nguyen2026ai} orchestrates response strategies using runtime state maintained by a digital twin, Gao et al.~\cite{gao2026context} continually revise response decisions through tool execution and environmental observations, and SecLoop~\cite{cao2025advancing} connects security-strategy generation, execution validation, and feedback optimization in a closed loop.

These studies provide different forms of state maintenance, stage organization, historical experience, and execution feedback.
For compromised software systems, decisions must track discovered attack relations, remaining adversarial capabilities, and prior response effects on these capabilities and service state.
States centered on fault localization and availability do not directly cover these requirements.
Security-response methods use stage-level recovery states, observation--action histories, digital-twin context, or strategy feedback, but stage progress and strategy evaluations alone do not represent attack entities and response relations.
The challenge is to continuously maintain these entities and relations across steps and stages~\cite{liu2024lost,zhang2025survey}, so that evidence, response outcomes, and unresolved conditions jointly support decisions and experience reuse under applicable conditions~\cite{zhao2024expel,shinn2023reflexion,xu2026mem,ding2012healing}.

To this end, we propose STAIR, an end-to-end agentic planning framework maintaining attack relations and response progress through \ac{gas}.
GAS associates affected entities, attack relations, response actions, and execution feedback in one incident state, keeping discovered but unresolved conditions visible across steps and stages.
The Planner determines the response stage and recovery focus from the latest GAS and retrieves effect-validated historical experience using the same state and stage.
Execution feedback updates GAS for subsequent response.

Realizing this insight requires addressing three challenges.
First, new evidence and response actions continually change incident understanding and environmental state, requiring state continuity across steps.
STAIR incrementally maintains affected entities, attack relations, response actions, and execution feedback in \ac{gas}, keeping discovered but unresolved conditions available to subsequent decisions.
Second, historical action reuse must account for applicable conditions and actual effects.
The Execution Harness validates effects through pre- and post-action environmental checks; the Experience Database binds each action to its pre-action state, response stage, and validated outcome, aggregating records under similar conditions to provide references to effective and high-risk actions.
Third, response actions must stay aligned with recovery objectives as these change with new evidence and execution results.
The Stage Router determines the current stage and recovery focus from the latest GAS, supports advancement and rollback, and dispatches the corresponding stage-specialized agent to organize working memory, experience, and tool use.

We implement a prototype supporting graph-based state management, experience retrieval, stage-specialized planning, and bounded tool execution over 33 response capabilities.
We construct STAIR-Bench, an executable incident response benchmark containing 44 Docker-based cyber-range cases, and evaluate overall effectiveness, performance under different incident complexities, and the contribution of core design mechanisms on held-out cases.
STAIR achieves an overall defense score of 0.95, improving over the strongest baseline by approximately 10.47\%.
We further evaluate dynamic response under ongoing attacks and benign activity in the independent CAGE4 environment~\cite{kiely2025exploring}; compared with the Claude Code baseline using the same model backend, STAIR reduces cumulative penalties by approximately 28.8\%.

In summary, this paper makes three main contributions.
\begin{itemize}
    \item We formulate security incident response planning for live software systems as continuous decision making that tracks attack relations, response progress, and unresolved conditions while restoring services, supporting decisions across steps and stages.

    \item We design STAIR, whose GAS continuously associates entities, attack relations, responses, and feedback to ground stage and recovery-focus selection and effect-validated experience reuse in an evolving attack and defense state.

    \item We construct the executable incident response benchmark STAIR-Bench and evaluate STAIR on this benchmark and the independent CAGE4 dynamic adversarial environment, using overall comparisons, complexity analysis, ablations, and representative traces to assess response effectiveness and the roles of its design mechanisms.
\end{itemize}

\section{Background and Related Work}
\label{sec:background}

\subsection{AIOps and Security Incident Response}
\label{subsec:background-ir}

AIOps uses data-driven and intelligent techniques to support the operation of software systems~\cite{zhang2025survey,chen2020towards,lou2017experience}.
Its scope has expanded from runtime monitoring, anomaly detection, and root-cause analysis toward automated remediation, incident handling, and next-action decision making~\cite{remil2024aiops,ahmed2023recommending}.
As this automation broadens, AIOps increasingly moves beyond identifying operational abnormalities to taking actions from runtime information and revising subsequent decisions as the system evolves~\cite{chen2026stratus,chen2025aiopslab,roy2024exploring}.

Security incident response is an important security-focused instance of this runtime incident-handling paradigm.
It is a multi-stage operational process for continuously mitigating cyberattacks and restoring affected software systems~\cite{scarfone2008sp, hammar2025incident, gao2026context}.
Standard guidance organizes incident handling into stages such as containment, assessment, preservation, eviction, hardening, and restoration~\cite{schlette2024you, stevens2022ready}.
These stages reflect different operational objectives, and response decisions change as the incident progresses and the runtime environment evolves~\cite{akbari2024requirements,shaked2023operations}.
Unlike ordinary fault recovery, restoring service availability does not imply that attacker capabilities have been removed: stolen credentials, persistence mechanisms, malicious artifacts, or exploitable paths may remain.
Meanwhile, response actions such as host isolation, process termination, artifact removal, and service restart can suppress attacker capabilities while disrupting legitimate services or destroying forensic evidence.

The decision basis for security incident response also changes throughout the response process.
Incident analysis draws on alerts, logs, and other defender-visible evidence, while estimates of the incident scope and affected assets are progressively refined as more information becomes available~\cite{nelson2025incident,alahmadi202299,yang2024true}; response actions themselves also change the current environment.
In this paper, the incident state captures attacker activities, affected assets, evidence, and previous response results, and evolves as the response proceeds.
Security incident response planning is therefore a continuous closed-loop decision process: responders select the next action based on the current incident state, observe environmental changes after execution, and adjust subsequent actions accordingly~\cite{hammar2025adaptive,gao2026context}.
Execution results indicate whether an operation completes, while its response effect further depends on whether it advances the current recovery objective and whether it introduces side effects such as service disruption or evidence loss.

\subsection{Automated Incident Handling}
\label{subsec:related-work}

Recent AIOps research has developed increasingly capable mechanisms for automated incident handling~\cite{chen2026stratus,jha2025itbench,chen2025aiopslab,ahmed2023recommending,roy2024exploring}.
E2E-REME~\cite{zhang2026e2e} connects microservice diagnosis with executable remediation and uses prior remediation outcomes to improve subsequent repairs.
Triangle~\cite{yu2025triangle} automates production incident triage through a multi-agent workflow that gathers troubleshooting information and supports mitigation.
GALA~\cite{tian2026gala} organizes runtime evidence through service-dependency graphs and shared investigation information to support diagnosis and response recommendations.
For security incidents, Hammar et al.~\cite{hammar2025incident} represent recovery state with six stage-completion flags and evaluate actions through predicted recovery trajectories.
SOAR4BC~\cite{nguyen2026ai} integrates digital-twin and security context in a knowledge base to generate and execute adaptation workflows from continuously updated observations.
Gao et al.~\cite{gao2026context} infer recovery state and attack models from observation--action histories, calibrating planning against actual observations.
SecLoop~\cite{cao2025advancing} optimizes security strategies using feedback on execution, attack outcomes, and service availability.

Table~\ref{tab:background_comparison} summarizes these capability differences.
Fault-oriented methods organize remediation around diagnostic evidence and repair outcomes; security-response methods organize decisions around recovery estimation, history-conditioned state inference, system adaptation, or strategy optimization.
STAIR continuously maintains incident entities and relations: GAS associates affected entities, attack relations, executed responses, and feedback, retaining discovered but unresolved conditions across steps and stages.
These conditions inform stage assessment and recovery-focus selection, while the same GAS and current stage condition historical action retrieval.

\begin{table}[t]
\centering
\footnotesize
\caption{Capability comparison of representative automated incident-handling approaches.}
\label{tab:background_comparison}
\renewcommand{\arraystretch}{1.08}
\setlength{\tabcolsep}{7pt}
\begin{tabular}{@{}lccc@{}}
\toprule
\textbf{Method}
& \textbf{Incident State}
& \textbf{Stage Planning}
& \textbf{Historical Experience} \\
\midrule

E2E-REME~\cite{zhang2026e2e}
& $\circ$ & \xmark & \checkmark \\

Triangle~\cite{yu2025triangle}
& $\circ$ & \xmark & \xmark \\

GALA~\cite{tian2026gala}
& $\circ$ & \xmark & \xmark \\

Hammar et al.~\cite{hammar2025incident}
& \checkmark & \checkmark & \xmark \\

SOAR4BC~\cite{nguyen2026ai}
& \checkmark & \xmark & $\circ$ \\

Gao et al.~\cite{gao2026context}
& \checkmark & \checkmark & \xmark \\

SecLoop~\cite{cao2025advancing}
& $\circ$ & \xmark & $\circ$ \\

\addlinespace[0.15em]
\textbf{STAIR}
& \checkmark & \checkmark & \checkmark \\

\bottomrule
\end{tabular}

\vspace{0.35em}
\begin{minipage}{0.86\columnwidth}
\scriptsize
\textbf{Legend.}
\checkmark: explicitly supported;
$\circ$: partially or implicitly supported;
\xmark: not a primary capability.
Incident State: explicit maintenance and update of evolving incident or recovery state.
Stage Planning: explicit use of response-stage progress or the current response stage to organize action selection.
Historical Experience: reuse of response experience from prior incidents or response trajectories.
\end{minipage}
\end{table}



\section{Overview}

\subsection{Insight and Challenges}
\label{sec:motivation}

Our key insight is to organize stage-aware response and experience reuse around a continuously maintained attack and defense state.
GAS relates attack conditions to response progress, retaining conditions that remain unresolved after local actions.
The Stage Router determines the current stage and recovery focus from the latest GAS, returning to an earlier stage when new evidence makes its objectives unresolved again.
Experience retrieval uses the same GAS and stage to reference validated effects of historical actions under relevant conditions.
Execution feedback continuously updates GAS, grounding the next response in attack conditions and actual response progress.

Based on this insight, we identify three key challenges.

\noindent\textbf{Challenge 1: Maintaining Incident-State Continuity.}
Incident response is a continuously evolving process.
New alerts, investigation results, and response actions repeatedly change the understanding of affected assets, attack paths, and recovery progress.
For LLM-based response systems that rely on textual contexts or interaction histories, important early evidence, prior actions, and unresolved objectives can be lost over long-running interactions~\cite{liu2024lost,zhang2025survey}.
The key challenge is how to maintain an incrementally updated incident state shared across response steps, so that subsequent stage assessment, experience retrieval, and response planning remain grounded in the current response progress.

\noindent\textbf{Challenge 2: Reliable Experience Reuse.}
Historical response trajectories contain many executed actions, but these actions cannot be directly treated as reliable experience for future incidents.
Tool execution success does not imply that a recovery objective was advanced~\cite{bridges2023testing}; the same action may produce different effects under different incident states~\cite{shaked2023operations}, and a single historical outcome may also be affected by incidental environmental factors.
The key challenge is how to associate historical actions with both the incident state in which they were executed and their actual response effects, and to form reliable experience from repeated historical executions so that future planning can reuse effective operations under similar conditions while avoiding actions that produced high-risk outcomes.

\noindent\textbf{Challenge 3: Maintaining Recovery-Objective Alignment.}
Even with a continuous incident state and reusable historical experience, long-horizon response planning must continually determine which recovery objective should be advanced next.
Different response stages have different objectives and operational constraints, and these objectives do not necessarily progress in a one-way sequence: new security alerts, environmental evidence, or execution feedback can make an earlier objective unresolved again.
The key challenge is how to dynamically determine the current recovery stage and concrete recovery focus from the evolving incident state, while supporting stage advancement, rollback, and re-entry so that each response action remains aligned with the current recovery progress.

\begin{figure*}[t]
    \centering
    \includegraphics[width=\textwidth]{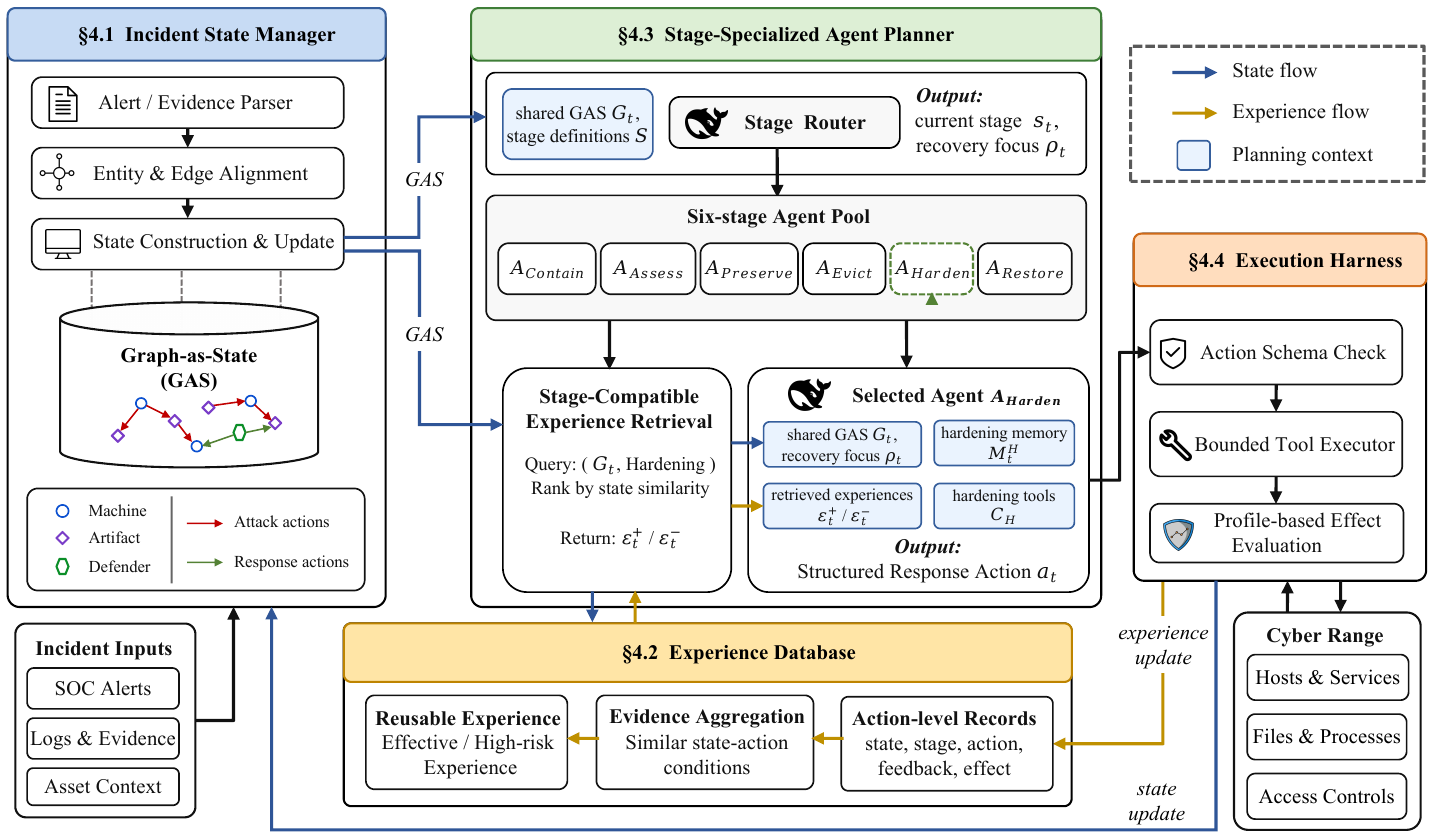}
    \caption{The workflow of the proposed agentic incident response framework.}
    \label{fig:architecture}
\end{figure*}

\subsection{Architecture Overview}
As shown in Fig.~\ref{fig:architecture}, the Incident State Manager continuously maintains the current incident as GAS; the Experience Database provides response experience grounded in historical state--action--effect records; the Stage-Specialized Agent Planner determines the current recovery stage and recovery focus from the latest GAS and generates the next response action using the corresponding experience and response capabilities; and the Execution Harness executes the action, returns feedback to update GAS, and validates action effects for future experience reuse.

\noindent\textbf{Incident State Manager (\S\ref{subsec:state-manager}).}
It maintains the shared incident state throughout the response process using GAS.
New security alerts, environmental evidence, response actions, and execution feedback are continuously incorporated into the graph, so subsequent stage assessment, experience retrieval, and response planning operate on the same up-to-date state.

\noindent\textbf{Experience Database (\S\ref{subsec:experience-db}).}
It organizes historical response actions as action-level experience by associating each action with its pre-action incident state, response stage, execution feedback, and validated effect, while aggregating repeated evidence under similar state--action conditions.
During the current response, it retrieves effective and high-risk experience according to the latest GAS and recovery stage, providing the Planner with reusable historical references and risk warnings.

\noindent\textbf{Stage-Specialized Agent Planner (\S\ref{subsec:planner}).}
It continuously determines the current recovery stage and concrete recovery focus from the latest GAS and dispatches planning to the corresponding stage-specialized agent.
The selected stage organizes its working memory, retrieved experience, and available tools, while all stage agents share the full GAS.
This allows planning to advance, roll back, or re-enter earlier stages as the incident state evolves, keeping response actions aligned with the current recovery objective.

\noindent\textbf{Execution Harness (\S\ref{subsec:execution-harness}).}
It maps structured response actions to concrete security tools or scripts and executes them in the target environment.
Real execution feedback updates GAS, while independently validated action effects are used to construct subsequent reusable experience.
The Harness therefore closes both the state-update loop and the experience-update loop.

\section{Design}
\label{sec:design}

\subsection{Incident State Manager}
\label{subsec:state-manager}

To address \textbf{Challenge 1}, we design the Incident State Manager to maintain a persistent and evolving incident state for response planning.
Specifically, we represent the incident as \ac{gas}, which unifies SOC alerts, environmental evidence, response actions, and execution feedback into a structured state representation.
Unlike textual contexts or interaction histories that require repeated reconstruction of incident progress, GAS continuously maintains the current response-relevant state as the incident evolves, providing a shared state foundation for subsequent stage assessment, experience retrieval, and response planning.

GAS represents the incident state \(G_t\) at response step \(t\) as an attributed graph:
\begin{equation}
G_t = (V_t, E_t, A_t),
\end{equation}
where \(V_t\) denotes response-relevant entities, \(E_t\) denotes behavioral relations formed during the incident, and \(A_t\) stores node- and edge-level attributes.
GAS jointly captures how attacker behaviors affect the environment and how defense actions change the current incident state.

Specifically, GAS models three categories of entities involved in incident response:
\begin{equation}
V_t = V_{t,M} \cup V_{t,A} \cup V_{t,D},
\end{equation}
where \(V_{t,M}\), \(V_{t,A}\), and \(V_{t,D}\) denote machine, artifact, and defender nodes, respectively.
Machine and artifact nodes represent affected runtime entities and incident-related objects, while defender nodes represent response actors that perform mitigation actions.

The edges in GAS describe the behavioral relations among these entities:
\begin{equation}
E_t = E_{t,atk} \cup E_{t,rsp},
\end{equation}
where \(E_{t,atk}\) captures attack behaviors reconstructed from SOC alerts and subsequent evidence, and \(E_{t,rsp}\) captures response actions performed during incident handling.
Each edge \(e=(u,v,\eta)\) records a concrete incident behavior, where \(u,v\in V_t\) denote the source and target nodes, and \(\eta\) stores attributes including behavior type, response stage, and contextual metadata.

The State Manager initializes the incident state \(G_0\) from the initial SOC alert.
During initialization, the system extracts response-relevant entities, behavioral relations, and contextual information from the alert and maps them to GAS nodes, edges, and attributes.
For example, affected hosts and attack-related artifacts are mapped to entity nodes, observed attack behaviors are mapped to attack relations, and alert sources and behavior attributes are retained as state information.

During response, the State Manager updates GAS after each response step.
At step \(t\), the Planner generates one response action \(a_t\) based on the current state \(G_t\), and the Execution Harness executes the action and returns execution feedback.
Newly arriving security alerts and environmental evidence are incorporated through the same update process.
The State Manager then merges these incremental inputs into \(G_t\): the executed action is appended as a new response edge, newly observed attack evidence supplements or extends attack-related relations, and execution feedback and environmental changes update the attributes of the corresponding entities and relations.
The resulting graph forms the next incident state \(G_{t+1}\).

To maintain consistency across incremental updates, the State Manager aligns newly arriving information with existing GAS nodes using deterministic entity identifiers.
For machine nodes, the identifier is constructed from the normalized hostname or IP address; for artifact nodes, it is jointly determined by the owning machine, artifact type, and normalized path.
If the generated identifier already exists, the system updates the corresponding node; else, a new node is created.
Each executed response action is appended as a distinct response edge and retains its execution order through the response stage and step index, while alert-derived evidence retains its source-log information.

\subsection{Experience Database}
\label{subsec:experience-db}

To address \textbf{Challenge 2}, we design the Experience Database to transform historical response executions into reusable action-level experience.
Execution success alone does not indicate whether an action actually advances the recovery objective or introduces additional risk~\cite{jha2025itbench,shaked2023operations}.
We therefore use each individual response action as the basic unit of experience and organize its incident state, response stage, execution feedback, and validated action effect into a structured experience record.

For each executed response action, we define an experience record as:
\begin{equation}
x_i=(G_i,s_i,a_i,y_i,f_i),
\end{equation}
where \(G_i\) denotes the incident state before the action, \(s_i\) denotes the corresponding response stage, \(a_i\) denotes the executed response action, \(y_i\) denotes the action effect validated by the Execution Harness, and \(f_i\) denotes the execution feedback summary.
The effect-validation process is described in \S\ref{subsec:execution-harness}.
Binding each action outcome to its pre-action state and response stage captures the conditions under which it was produced, enabling retrieval of stage-compatible actions executed under similar incident conditions.

A single historical execution may be affected by incidental environmental factors and should not by itself define a stable response pattern.
The Experience Database therefore aggregates repeated historical evidence under similar state--action conditions.
Records are first grouped by response stage, response operation, and target-object type, and records with sufficiently similar pre-action states are then organized into the same experience cluster using the same GAS similarity measure employed during retrieval.
Each cluster retains its underlying atomic records, summarizes the occurrences of \textsc{Effective}, \textsc{HighRisk}, and \textsc{Ineffective} outcomes, and keeps representative effective and high-risk records when available.
Thus, repeated executions under similar conditions are preserved as historical evidence rather than collapsed into a single label; ineffective outcomes remain in the cluster statistics but are not directly exposed to the Planner.

During response planning, the Experience Database retrieves action-level records using the current incident state \(G_t\) and response stage \(s_t\).
Records are filtered by stage and effect label, then ranked by GAS similarity, historical recovery progress or risk, and attack-type matching.
Effective and high-risk records are retrieved separately to form the positive experience set \(\mathcal{E}_t^+\) and negative experience set \(\mathcal{E}_t^-\), respectively.
The positive set provides response references that were effective under similar incident conditions, while the negative set warns against actions that produced high-risk outcomes.
Both sets are supplied to the selected stage agent.

\subsection{Stage-Specialized Agent Planner}
\label{subsec:planner}

To address \textbf{Challenge 3}, we design the Stage-Specialized Agent Planner to organize long-horizon incident response as a stage-aware planning process.
At each response step, the Stage Router determines the recovery stage and concrete recovery focus to advance from the latest GAS, and dispatches planning to the corresponding stage-specialized agent.
The current stage organizes stage-specific working memory, experience retrieval, and tool access, while GAS remains the shared incident state across stages.

\noindent\textbf{Stage Router.}
Following the incident response lifecycle described in \S\ref{sec:background}, we define the ordered recovery-stage sequence
\begin{equation}
\mathcal{S}=(\textsc{Containment}, \textsc{Assessment}, \textsc{Preservation}, \textsc{Eviction}, \textsc{Hardening}, \textsc{Restoration}).
\end{equation}
The order in \(\mathcal{S}\) defines the routing precedence.
\textsc{Containment} blocks attacker spread and control; \textsc{Assessment} confirms impact scope and affected assets; \textsc{Preservation} protects forensic evidence; \textsc{Eviction} removes attacker footholds; \textsc{Hardening} fixes vulnerable entry points and configuration weaknesses; and \textsc{Restoration} restores service availability and user access.

An LLM assesses whether each stage objective is satisfied and identifies unresolved recovery needs from the latest GAS; the Stage Router then selects the first stage requiring further progress according to the precedence in \(\mathcal{S}\).
If the objective of the current stage is satisfied, the Router advances to the next stage that still requires progress; if new security alerts, environmental evidence, or execution feedback make an earlier-stage objective unresolved again, the Router returns to that earlier stage.
This check is repeated at every planning step using the latest GAS.
At response step \(t\), the Router outputs \(z_t=(s_t,\rho_t)\), where \(s_t\in\mathcal{S}\) is the selected recovery stage and \(\rho_t\) is its concrete recovery focus, such as an uncontained attack path, an evidence object to preserve, a persistence item to remove, or a service to restore.

We denote the stage-specialized agents as \(\mathcal{A}=\{A_{\sigma}\mid\sigma\in\mathcal{S}\}\), where \(A_{\sigma}\) is responsible for planning under stage \(\sigma\).
Given \(s_t\), the Planner invokes \(A_{s_t}\) and uses the selected stage to organize its working memory, historical experience, and available tools.

\noindent\textbf{Stage-Specialized Planning.}
The selected agent \(A_{s_t}\) generates the next response action from the full GAS \(G_t\) and recovery focus \(\rho_t\).
All stage agents share the same GAS, so attack evidence, prior response actions, and recovery state established in earlier stages remain available to subsequent planning.
Stage specialization is realized through stage-specific planning history, historical experience, and bounded response capabilities.

The stage working memory preserves within-stage planning continuity.
For stage \(\sigma\), we define
\begin{equation}
M_t^{\sigma}=\{(a_\tau,f_\tau)\mid s_\tau=\sigma,\tau<t\},
\end{equation}
where \(a_\tau\) is a previous response action in stage \(\sigma\), and \(f_\tau\) is the corresponding execution feedback.
When \(A_{\sigma}\) is invoked again, these prior actions and feedback are included in its planning context, allowing the agent to account for previous attempts, failed outcomes, and within-stage progress.

Historical experience is also organized by the current stage.
The Experience Database first filters stage-compatible records using \(s_t\), and then retrieves positive and high-risk experiences according to the current GAS.
Each stage agent is configured with stage-specific experience-use instructions that guide how these retrieved experiences should be used under the current response objective, such as emphasizing evidence integrity during \textsc{Preservation} or service continuity during \textsc{Restoration}.

Each stage is also associated with a bounded tool subset \(\mathcal{C}_{\sigma}\subseteq\mathcal{C}\) aligned with its response objective.
These include diagnostic tools for \textsc{Assessment}, evidence-preservation tools for \textsc{Preservation}, isolation and access-control tools for \textsc{Containment}, cleanup tools for \textsc{Eviction}, patching and configuration tools for \textsc{Hardening}, and service recovery and validation tools for \textsc{Restoration}.

Based on this stage-conditioned context, \(A_{s_t}\) outputs a structured response action \(a_t=(T_t,P_t,R_t)\), where \(T_t\in\mathcal{C}_{s_t}\) is the selected response capability, \(P_t\) contains the arguments satisfying the tool schema, and \(R_t\) records the rationale.
The action is then passed to the Execution Harness for validation and execution, and the returned feedback updates GAS for the next planning step.

\subsection{Execution Harness}
\label{subsec:execution-harness}

The Execution Harness maps structured response actions generated by the Planner to the execution environment and feeds real execution outcomes back into the incident-response loop.
Given \(a_t=(T_t,P_t,R_t)\), the Harness first validates the selected response capability and its arguments against the corresponding tool schema, grounds the action to a concrete security tool or response script according to the tool catalog, and executes it in the target environment.
The system executes one response action at a time and returns unified execution feedback \(f_t\), including execution status, tool outputs, and newly observed environmental evidence.

Execution outcomes first update the incident state.
The Harness records the executed action as a new response-action edge and passes its execution feedback and newly observed environmental evidence to the Incident State Manager.
Following the incremental update mechanism in \S\ref{subsec:state-manager}, the State Manager updates the related entities and relations to produce the next incident state \(G_{t+1}\).

Beyond recording whether an action executes successfully, the Harness independently validates its actual response effect.
Each incident is associated with an evaluation profile that specifies executable validation checks for its recovery objectives together with policy and service-continuity constraints.
For action \(a_t\), the Harness records the relevant check states before execution, reruns the corresponding checks after execution, and derives an action-effect label
\begin{equation}
y_t \in \{\textsc{Effective},\textsc{HighRisk},\textsc{Ineffective}\}.
\end{equation}
These checks observe runtime states related to the response objective, such as whether a malicious process or persistence artifact remains present, whether a target port or service has been restored, and whether the response action violates required service continuity.
The execution feedback \(f_t\) therefore describes how the action executed and what the tool returned, whereas \(y_t\) captures whether the action actually advanced recovery or introduced unacceptable risk.

The evaluation profile is used only by the Execution Harness for effect validation; it is neither exposed to the Planner nor written into GAS.
After validation, the Harness passes the pre-action incident state \(G_t\), response stage \(s_t\), executed action \(a_t\), validated effect \(y_t\), and execution feedback \(f_t\) to the Experience Database, forming the action-level record defined in \S\ref{subsec:experience-db}.
In this way, the Harness closes two feedback paths: real execution results update the current GAS, while independently validated action effects provide the basis for subsequent experience reuse.

\subsection{Running Example}

\begin{figure}[t]
    \centering
    \includegraphics[width=0.75\linewidth]{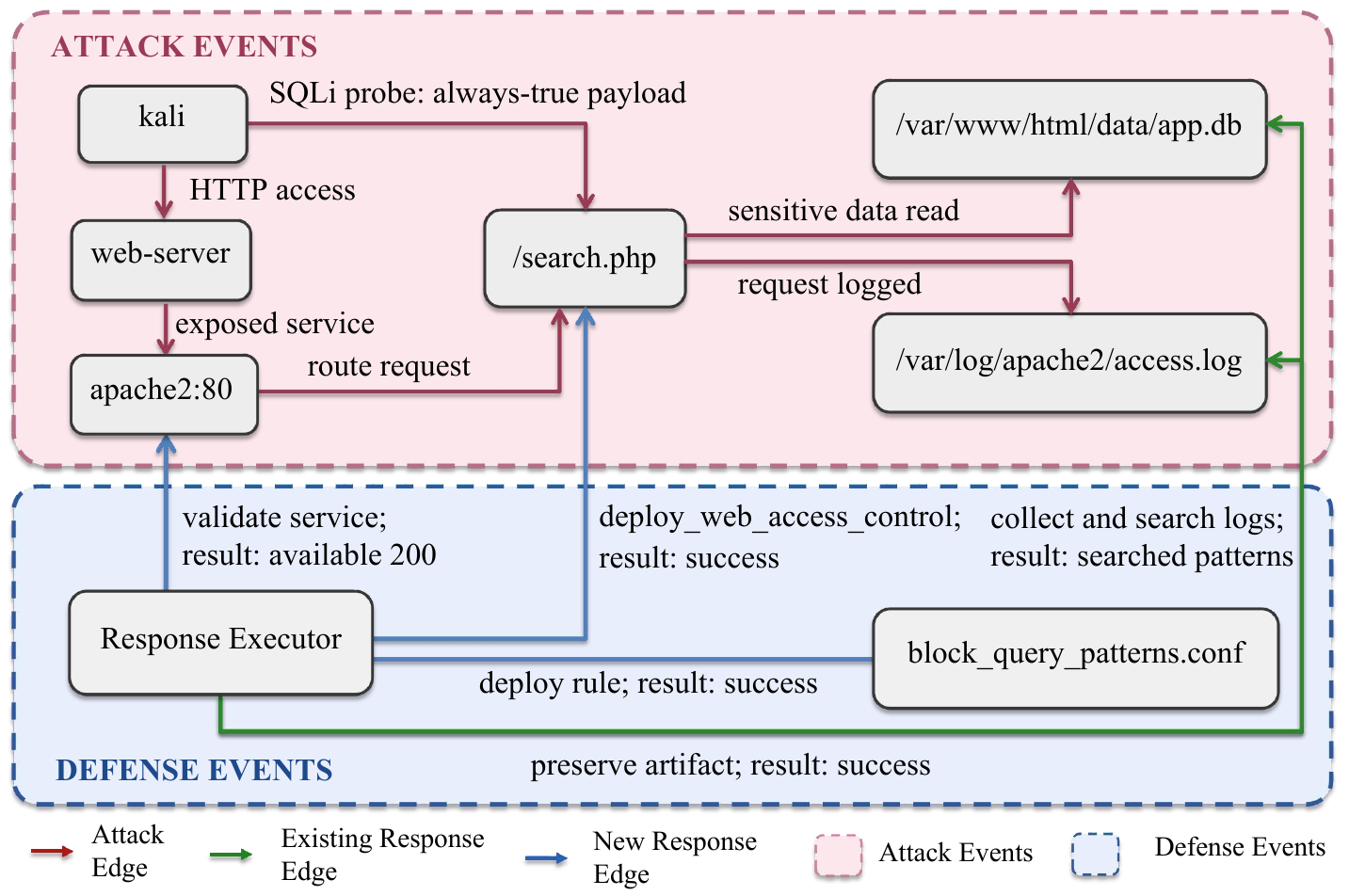}
    \caption{GAS snapshot and hardening update in the running example.}
    \label{fig:running-gas-update}
\end{figure}

We use a Web SQL injection incident to illustrate one closed-loop response step grounded in GAS, stage-specialized planning, tool execution, and experience update. 
The incident involves an attacker host \texttt{kali} and a victim host \texttt{web-server}. 
The attacker probes the public \texttt{/search.php} endpoint with an always-true payload and replays a \texttt{UNION SELECT} query to read sensitive records from \texttt{app.db}. 
Figure~\ref{fig:running-gas-update} shows the GAS snapshot around the hardening step, including the reconstructed attack chain, prior response actions, and the new hardening response edges.

At this step, the Stage Router selects \textsc{Hardening} as the stage to advance, with the recovery focus of blocking the observed SQL injection replay. 
The hardening agent plans with the shared GAS, stage working memory, retrieved experiences, and hardening tools. 
The positive experience suggests adapting a query-blocking pattern, while the negative experience warns against shutting down the Web service. 
Figure~\ref{fig:running-prompt} shows the action-generation prompt formed from the recovery stage, recovery focus, shared GAS and stage memory, retrieved experiences, hardening tool set, and output schema.

Given this prompt, the hardening agent follows its stage-specific experience-use instructions to adapt the blocking pattern to the affected endpoint and outputs \texttt{deploy\_web\_access\_control}. 
The Execution Harness validates the action, grounds it into the corresponding response script, and executes it on \texttt{web-server}. 
As shown in Figure~\ref{fig:running-gas-update}, execution feedback is written back to GAS as response-action edges from the defender node to the affected endpoint, configuration artifact, and service. 
The related artifact state is updated by marking \texttt{block\_query\_patterns.conf} as present.

The Harness also validates the action effect using the incident-specific evaluation profile. 
It confirms that the replay is blocked, the blocking rule is present, and the Web service remains reachable, and therefore labels the action as \textsc{Effective}. 
The pre-action GAS, response stage, executed action, effect label, and feedback summary form the action-level experience record, which is then incorporated into the Experience Database for future reuse.

\begin{figure}[t]
\centering
\begin{minipage}{0.95\linewidth}
\begin{promptbox}{Compressed Action Generation Prompt}
\scriptsize

\textbf{\#\#\# Current recovery stage \(s_t\):} \\
\textsc{Hardening}

\vspace{1mm}
\textbf{\#\#\# Recovery focus \(\rho_t\):} \\
Block the observed SQL injection replay against the search endpoint while preserving Web service availability.

\vspace{1mm}
\textbf{\#\#\# Planning context:} \\
\textbf{Shared GAS.}
kali \(\rightarrow\) /search.php: SQLi probe/replay;
 /search.php \(\rightarrow\) app.db: sensitive data read;
 access.log records UNION SELECT payload;
 previous responses collected and preserved access.log/app.db. 

\vspace{0.5mm}
\textbf{Stage memory.}
Prior hardening attempts: None.

\vspace{1mm}
\textbf{\#\#\# Retrieved experiences:} \\
\textbf{Positive \(\mathcal{E}_t^+\).}
SQLi query blocking: \textsc{Effective}; blocking malicious query patterns stopped replay of UNION SELECT payloads. 

\vspace{0.5mm}
\textbf{Negative \(\mathcal{E}_t^-\).}
Service shutdown: \textsc{HighRisk}; stopping the Web service removed the attack surface but violated service continuity.

\vspace{1mm}
\textbf{\#\#\# Hardening tool set \(\mathcal{C}_{H}\):} \\
Available operation: deploy\_web\_access\_control. \\
Required arguments: block patterns, test query strings, execution target. \\
Available operation: ...

\vspace{1mm}
\textbf{\#\#\# Output schema:} \\
Return one action with tool, operation, arguments, and reason.

\end{promptbox}
\end{minipage}

\caption{Action-generation prompt for the hardening step in the running example.}
\label{fig:running-prompt}
\end{figure}
\section{Implementation}

The main pipeline comprises about 21.03 KLOC for graph-based state management, experience construction and retrieval, stage-specialized LLM planning, and bounded tool execution.

\noindent (a) \ac{gas} is implemented as a graph-state runtime that maintains nodes, edges, and attributes for the evolving incident state.
Security alerts, environmental evidence, and execution feedback are incrementally incorporated into the graph.
Machines are aligned by normalized hostname or IP, while artifacts are aligned by their owning machine, artifact type, and normalized path; executed response actions are appended as distinct response edges.
Stage agents use the same continuously updated full GAS for planning, while stage specialization is introduced through routing, working memory, retrieved experience, and tool access rather than stage-specific state projection.

\noindent (b) The Experience Database stores action-level records containing the pre-action GAS, response stage, executed action, validated effect label, and execution feedback. In our prototype, it is constructed offline from historical response trajectories in the training environments and remains fixed during testing. Records are organized by response stage, response operation, and target-object type. Serialized GAS states are embedded using \texttt{stella\_en\_1.5B\_v5}. Retrieval filters candidates by stage and effect label, then ranks them using cosine similarity, recovery progress or risk scores, and attack-type matching. Candidates with raw cosine similarity below 0.05 are excluded, and up to three effective records and three high-risk records are retrieved separately.

\noindent (c) The Stage-Specialized Agent Planner uses DeepSeek V4 Pro as the default LLM backend.
It implements the Stage Router, six stage-specialized agents, stage working memories, stage-specific experience-use instructions, and stage tool subsets.
At each planning step, the Router uses LLM assessments of the latest GAS to select the first unresolved stage under the predefined precedence, supporting advancement and rollback.
The selected agent generates a structured response action from the full GAS, current recovery focus, stage working memory, retrieved positive and high-risk experience, and the tools available to that stage.

\noindent (d) The Execution Harness exposes 33 bounded response capabilities covering log and artifact collection, host and service inspection, network and firewall control, cleanup, and service restart.
It validates generated actions against tool schemas, grounds them to concrete tools or scripts, executes them in Docker 24.0.7 and containerlab 0.74.3 cyber ranges, and returns unified execution feedback for GAS updates.
Each incident also provides Harness-only executable validation checks over file, process, port, and service states.
The Harness compares the relevant checks before and after each action to derive \textsc{Effective}, \textsc{HighRisk}, or \textsc{Ineffective} effect labels.
The evaluation profile is not exposed to the Planner or written into GAS, and the validated effect together with execution feedback is passed to the Experience Database.

\section{Evaluation}
\label{sec:evaluation}

We evaluate STAIR from four perspectives: overall response effectiveness, robustness under incident complexity, response capability under dynamic adversarial conditions, and the contribution of its core design mechanisms. We focus on four research questions:

\begin{itemize}[noitemsep, topsep=1pt, partopsep=1pt, listparindent=\parindent, leftmargin=*]
    \item \textbf{RQ1: End-to-End Effectiveness.} Does STAIR improve overall end-to-end incident response effectiveness?
    \item \textbf{RQ2: Robustness under Incident Complexity.} Does STAIR maintain stable response quality as incident complexity increases?
    \item \textbf{RQ3: Dynamic Adversarial Response.} Can STAIR maintain effective closed-loop response in continuously evolving adversarial environments?
    \item \textbf{RQ4: Design Attribution.} How do STAIR's core design mechanisms affect response behavior and overall effectiveness?
\end{itemize}

\subsection{Experimental Setup}
\label{subsec:experimental-setup}

\noindent\textbf{STAIR-Bench.}
Existing public cyber-defense benchmarks are limited in their coverage of end-to-end cyber-range evaluation across diverse attack scenarios~\cite{standen2021cyborg,kiely2025exploring,bridges2023testing}. We therefore build STAIR-Bench, an executable benchmark for end-to-end security incident response based on Docker~\cite{docker} and Containerlab~\cite{containerlab}. It contains 44 cases spanning 14 attack families, 10 MITRE ATT\&CK~\cite{strom2018mitre} tactics, and 33 ATT\&CK techniques. Each case deploys runnable software services and executable attacks, and provides a cyber-range topology, an attack playbook, defender-visible telemetry, a response-action schema, and a hidden evaluation profile. Each trial starts from a clean topology snapshot and executes the case-specific attack playbook to establish the compromised environment before incident response begins.

\noindent\textbf{Train/test split.}
We explicitly organize the STAIR-Bench split by attack family. Of the 44 cases, 30 are used for framework calibration and Experience Database construction, 12 are same-family held-out cases, and 2 are OOD held-out cases. Same-family cases come from attack families represented in training but are independent incident instances, whereas OOD cases come from attack families absent from the training set. Test trajectories are not written back to the Experience Database. The split therefore evaluates response to both new instances of observed attack families and previously unseen attack families.

\noindent\textbf{Complexity grouping.}
For RQ2, we divide the held-out STAIR-Bench cases into Low, Medium, and High complexity tiers according to the number of attack vectors, involved protocols or services, and required cross-host and cross-objective coordination. These labels are used only for analysis and are hidden from the evaluated methods.

\noindent\textbf{CAGE4 environment.}
We additionally evaluate STAIR in CAGE4~\cite{kiely2025exploring}, which provides a dynamic enterprise environment with ongoing adversarial and benign activities.
CAGE4 runs Red, Blue, and Green agents in a randomized enterprise environment. Its FiniteStateRedAgent continues adversarial activity as the interaction unfolds, while Enterprise Green agents generate benign local-work and service-access activity. As a result, adversarial behavior, benign activity, and defender observations evolve jointly during response. We use 50 paired seeds, with 500 timesteps per episode.

\noindent\textbf{Test protocol.}
Across both benchmarks, evaluated methods follow the same defender information boundary. They can access only authorized topology or asset information, defender-visible alerts or telemetry, observations returned by response tools, and action-execution feedback. Attack playbooks, hidden attack-state information, attacker internal state, and evaluator-only information used for scenario construction or evaluation are never exposed to the planning context. Methods evaluated under the same setting use the same observation boundary and response capabilities.
On STAIR-Bench, all methods have a 30-minute wall-clock limit per trial.
CAGE4 uses paired seeds to control the initial scenario realization and the shared decision budget specified in RQ3.

\subsection{Baselines}
\label{subsec:baselines}

\noindent\textbf{Baseline selection.}
We compare with three representative baselines adapted to STAIR-Bench: Hammar et al., SecLoop, and Claude Code Agent, covering incident-response planning, reward-driven automation, and open-ended agentic interaction.

\noindent\textbf{Claude Code Agent.}
We use Claude Code as the open-ended agentic baseline~\cite{Anthropic}. 
It interacts with the live environment through the same response tools and continues planning from execution feedback. 
We instantiate it with Claude Opus 4.7, DeepSeek V4 Pro, and Kimi K3, denoted CC-Opus, CC-DeepSeek, and CC-Kimi, respectively. 
It does not use GAS, the Experience Database, or the Stage-Specialized Agent Planner.

\noindent\textbf{Hammar et al.}
We implement the incident-response planning method of Hammar et al.~\cite{hammar2025incident} using the paper's fine-tuned Qwen2.5-14B planner. 
Since it outputs textual response actions, we use an Action Router backed by DeepSeek V4 Pro to map them to our structured response capabilities. 
Unmapped or invalid actions are treated as unexecutable. 
The router only performs schema grounding and does not access hidden profiles or execution feedback.

\noindent\textbf{SecLoop.}
We implement SecLoop following its reward-driven execution-grounded setting~\cite{cao2025advancing}, using Qwen2.5-7B-Instruct as the base model. 
Since no trained checkpoint is released, we reproduce its training process on the STAIR-Bench training ranges. 
Evaluation is conducted only on held-out STAIR-Bench ranges, and test trajectories are not used for training.

\begin{table*}[t]
\centering
\caption{End-to-end incident response performance on held-out STAIR-Bench cases. Values are rounded to two decimal places; relative gains use the displayed values.}
\label{tab:main_results}
\footnotesize
\setlength{\tabcolsep}{2.0pt}
\begin{tabular*}{\textwidth}{@{\extracolsep{\fill}}lcccccccc@{}}
\toprule
\textbf{Method} &
\textbf{ODS$\uparrow$} &
\textbf{Recovery$\uparrow$} &
\textbf{AttackSup$\uparrow$} &
\textbf{AESR$\uparrow$} &
\textbf{SCS$\uparrow$} &
\textbf{ReattackBlk$\uparrow$} &
\textbf{TTR$\downarrow$} &
\textbf{ActProg$\uparrow$} \\
\midrule
STAIR &
\textbf{0.95} &
\textbf{0.91} &
\textbf{1.00} &
\textbf{0.92} &
\textbf{1.00} &
\textbf{1.00} &
18.57 &
\textbf{0.34} \\
\texttt{CC-DeepSeek} &
0.86 &
0.78 &
0.94 &
0.83 &
0.94 &
0.94 &
19.18 &
0.14 \\
\texttt{CC-Opus} &
0.80 &
0.70 &
0.88 &
0.82 &
0.91 &
0.88 &
22.00 &
0.13 \\
\texttt{CC-Kimi} &
0.78 &
0.67 &
0.79 &
0.86 &
\textbf{1.00} &
0.89 &
17.07 &
0.14 \\
Hammar et al. &
0.54 &
0.48 &
0.36 &
0.77 &
0.83 &
0.46 &
12.36 &
0.26 \\
\texttt{SecLoop} &
0.19 &
0.04 &
0.00 &
0.18 &
\textbf{1.00} &
0.11 &
\textbf{1.00} &
0.20 \\
\bottomrule
\end{tabular*}
\end{table*}

\subsection{RQ1: End-to-End Effectiveness}
\label{subsec:rq1}

\noindent\textbf{Protocol and metrics.}
RQ1 evaluates end-to-end incident response on held-out STAIR-Bench cases.
We run each method 10 times per test case, average the repeated runs within each case, and report macro-averaged results across test cases.
Recovery measures final satisfaction of remediation and restoration checks.
AttackSup measures suppression of attacker objectives, while ReattackBlk measures whether post-response attack replay is blocked.
AESR is the fraction of successfully executed actions, and ActProg is the fraction of actions making verified recovery progress, assessed through pre- and post-action checks.
SCS measures satisfaction of service-continuity requirements, and CDR measures response-induced collateral damage to protected assets or legitimate services.
TTR counts response steps until termination, including runs with incomplete recovery.
We use Overall Defense Score (ODS) as the primary composite metric:
\begin{equation}
\mathrm{ODS} = 0.40 \cdot \mathrm{Recovery} + 0.30 \cdot \mathrm{AttackSup} + 0.15 \cdot \mathrm{AESR} + 0.15 \cdot \mathrm{SCS}.
\end{equation}
Informed by NIST SP 800-61 Rev.~3 and CAGE4, which jointly motivate effective incident mitigation, recovery, and preservation of legitimate operations~\cite{nelson2025incident,kiely2025exploring}, we prioritize Recovery and AttackSup while retaining executability and service continuity.

\noindent\textbf{Results.}
Table~\ref{tab:main_results} shows that STAIR achieves the highest overall response quality.
Compared with the strongest baseline, CC-DeepSeek, STAIR improves ODS from 0.86 to 0.95 and Recovery from 0.78 to 0.91, with approximate relative gains of 10.47\% and 16.67\%, respectively.
AttackSup, ReattackBlk, and SCS reach 1.00.
Under the evaluated incidents and replay conditions, STAIR therefore achieves more complete recovery and attack suppression while preserving service continuity.
Both methods use DeepSeek V4 Pro, showing the overall benefit of the response framework with the same model backend.

The Claude Code backends further expose a gap between action execution and recovery completeness.
CC-Kimi has higher AESR (0.86) and SCS (1.00) than CC-DeepSeek and a shorter TTR (17.07 versus 19.18), but lower Recovery (0.67) and AttackSup (0.79), yielding an ODS of 0.78.
Thus, executable actions, limited service impact, and shorter trajectories do not ensure complete incident resolution.
Subsequent decisions must continue to address unrepaired system states and remaining attacker capabilities after local actions succeed.

These results are consistent with STAIR's mechanisms for retaining affected assets and unresolved conditions in GAS, prioritizing unmet stage objectives, and guiding actions with effect-validated experience.

Hammar et al. and SecLoop achieve ODS of 0.54 and 0.19, respectively, both below the Claude Code baselines.
In our adaptation, Hammar et al.'s textual plans lack execution-feedback-driven revision, limiting their ability to adjust subsequent actions to observed outcomes.
In our reproduction, SecLoop's response trajectories remain focused on assessment-like actions and leave attacker footholds and vulnerable entry points insufficiently remediated, with Recovery of 0.04 and AttackSup of 0.00.
These results indicate that long-horizon response requires both continuous plan revision from execution feedback and action selection aligned with actual recovery objectives.

\noindent\textbf{Weight sensitivity.}
STAIR is no worse than any baseline on each of the four ODS components and thus achieves the highest ODS under any nonnegative component weighting summing to one.

\begin{figure}[t]
\centering
\includegraphics[width=0.98\linewidth]{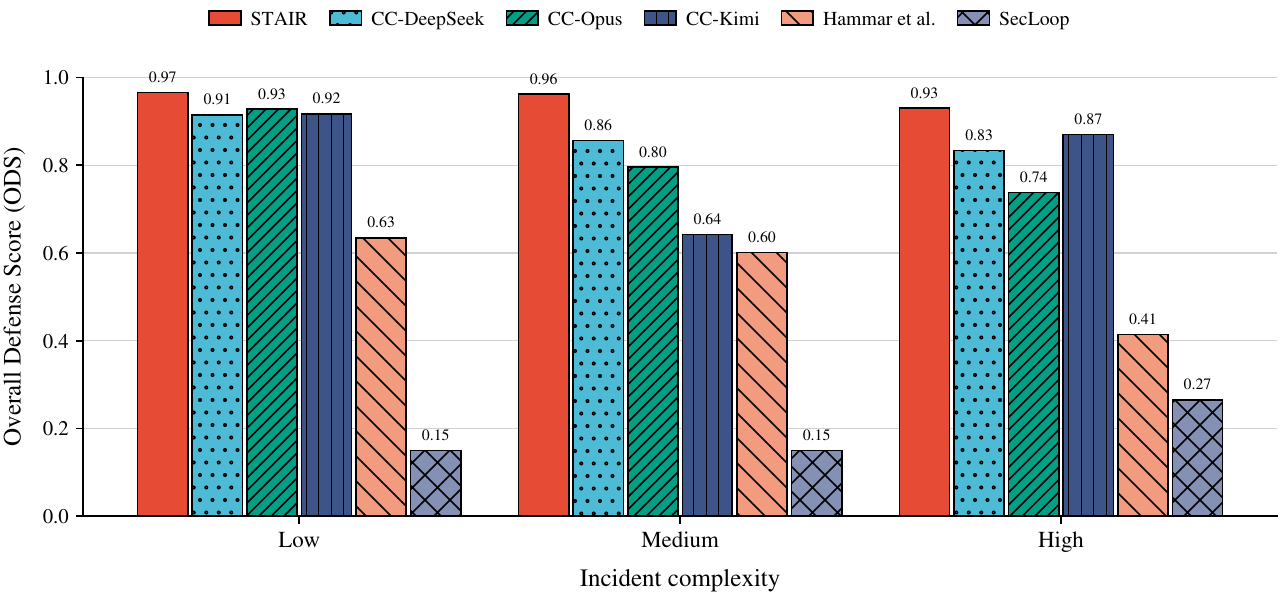}
\caption{Overall Defense Score (ODS) across incident complexity tiers on held-out STAIR-Bench cases.}
\label{fig:attack_complexity}
\end{figure}

\subsection{RQ2: Robustness under Incident Complexity}
\label{subsec:rq2}

\noindent\textbf{Protocol and metrics.}
RQ2 evaluates whether STAIR maintains stable response quality as incident complexity increases.
We group the 14 held-out STAIR-Bench cases into Low (3 cases), Medium (6 cases), and High (5 cases) tiers by structural characteristics, including attack vectors, involved protocols or services, and required cross-host or cross-stage state reconciliation.
We compare ODS across the three tiers to measure overall defense quality, and further use ActProg and execution traces to analyze response behavior under increasing complexity.

\noindent\textbf{Results.}
Figure~\ref{fig:attack_complexity} shows that STAIR achieves the highest ODS in all three complexity tiers, with scores of 0.97, 0.96, and 0.93 for Low, Medium, and High, respectively.
Using the displayed values, ODS decreases from Low to High by 4.12\% for STAIR, compared with 8.79\% for CC-DeepSeek, 20.43\% for CC-Opus, and 34.92\% for Hammar et al.
CC-Kimi shows a larger degradation in the Medium tier, mainly due to SMTP propagation cases in which it continues local remediation but does not fully eliminate the propagation source and related attack conditions, consistent with the gap between local action quality and overall recovery coverage observed in RQ1.
Overall, STAIR maintains more stable end-to-end defense quality as incident complexity increases.

\noindent\textbf{Analysis.}
STAIR achieves ActProg values of 0.30, 0.34, and 0.37 in the Low, Medium, and High tiers, whereas CC-DeepSeek achieves 0.14, 0.12, and 0.15, respectively.
STAIR consistently exceeds CC-DeepSeek with the same model backend, reaching a ratio of approximately 2.47 in the High tier.
Together with the tier-level ODS results, this shows that STAIR converts a larger fraction of actions into verified recovery progress while maintaining high response quality in structurally complex incidents typically requiring longer response horizons.

The \texttt{sh\_01} trace illustrates the mechanism behind this difference: CC-DeepSeek successfully deletes a marker created by attack replay, but does not subsequently address the FTP upload path that can recreate it, so the marker reappears after reattack.
In STAIR, the attack path and its unresolved condition remain in GAS after the local cleanup, so the Stage Router continues to organize subsequent planning around the incomplete recovery objective rather than allowing it to disappear from the decision context.
This trace shows how GAS-based state continuity and stage-aware planning help preserve unresolved conditions across longer response chains, consistent with STAIR's higher ActProg and more stable ODS under increasing incident complexity.

\subsection{RQ3: Dynamic Adversarial Response}
\label{subsec:rq3}

\noindent\textbf{Protocol and metrics.}
RQ3 evaluates dynamic response in CAGE4 (Section~\ref{subsec:experimental-setup}), comparing STAIR, CC-DeepSeek, Random, and No defense over 50 paired seeds with 500 timesteps per episode. STAIR, CC-DeepSeek, and Random share the same observation boundary and action space, and make decisions at initialization, upon new alerts or action completion, and during periodic checks, with at most 25 decisions per Blue agent. STAIR incorporates monitoring events, investigation evidence, and action feedback into GAS, and maps operations to native investigation, removal, and restoration actions. Since CAGE4 instances come from seed-based randomization of the same scenario generator, we omit scenario-specific historical experience for a fair comparison.

We use native cumulative reward, aggregated following the official evaluator, and report mean, standard deviation, 95\% confidence intervals (CIs), paired reward gains, and win rates.
We also report cumulative penalty reduction relative to No defense, computed as the increase in mean cumulative reward over No defense divided by the absolute value of No defense's mean cumulative reward, expressed as a percentage.

\begin{table}[t]
\centering
\caption{CAGE4 results over 50 paired seeds. Reward: mean \(\pm\) standard deviation; CI: 95\% confidence interval; reduction: cumulative penalty reduction relative to No defense.}
\label{tab:cage4_results}
\small
\setlength{\tabcolsep}{8.0pt}
\begin{tabular}{@{}lccc@{}}
\toprule
\textbf{Method} & \textbf{Reward$\uparrow$} & \textbf{95\% CI} & \textbf{Reduction$\uparrow$} \\
\midrule
STAIR & \(\mathbf{-4313 \pm 1204}\) & \(\mathbf{[-4641,-3983]}\) & \textbf{36.6\%} \\
\texttt{CC-DeepSeek} & \(-6061 \pm 1394\) & \([-6417,-5677]\) & 10.8\% \\
Random & \(-6776 \pm 1295\) & \([-7143,-6432]\) & 0.3\% \\
No defense & \(-6798 \pm 1108\) & \([-7107,-6498]\) & 0.0\% \\
\bottomrule
\end{tabular}
\end{table}

\noindent\textbf{Results.}
Table~\ref{tab:cage4_results} shows that STAIR achieves the highest mean cumulative reward, reducing cumulative penalties by approximately 36.6\% relative to No defense, compared with 10.8\% for CC-DeepSeek and 0.3\% for Random.
Compared with CC-DeepSeek using the same model backend, STAIR further reduces cumulative penalties by approximately 28.8\%.

Paired comparisons show that STAIR's mean reward gain over CC-DeepSeek is 1748, with a 95\% CI of [1324, 2166], and it achieves a higher reward on 42 of the 50 seeds (84\%).
The mean gains over Random and No defense are 2463 and 2486, with 95\% CIs of [2016, 2911] and [2162, 2819], and win rates of 90\% and 100\%, respectively.
All three paired-difference CIs exclude zero, and STAIR wins on most paired seeds, showing consistency across seeds.

\noindent\textbf{Analysis.}
Under the same decision-trigger and budget constraints, Random's mean cumulative reward is close to No defense, whereas STAIR and CC-DeepSeek achieve higher rewards.
This indicates that merely providing response opportunities is insufficient for effective defense in this setting; selecting actions based on observations and feedback also matters.
STAIR further outperforms CC-DeepSeek with the same model backend, showing that its overall response framework remains beneficial under continuous adversarial activity.
Unlike the post-response attack replay in RQ1, attacks and benign activity in CAGE4 continue during response; these results therefore provide complementary evidence of STAIR's effective dynamic response in an independent environment under the current decision budget.

\begin{table*}[t]
\centering
\caption{Cumulative component attribution on held-out STAIR-Bench ranges. G, P, and E denote GAS, the Stage-Specialized Agent Planner, and the Experience Database. AESR measures action execution success. CDR (Collateral Damage Rate) measures response-induced collateral damage to protected legitimate assets or services.}
\label{tab:ablation}
\footnotesize
\setlength{\tabcolsep}{2.00pt}
\begin{tabular*}{\textwidth}{@{\extracolsep{\fill}}lccccccccc@{}}
\toprule
\textbf{Method} &
\textbf{ODS$\uparrow$} &
\textbf{Recovery$\uparrow$} &
\textbf{AttackSup$\uparrow$} &
\textbf{AESR$\uparrow$} &
\textbf{SCS$\uparrow$} &
\textbf{ReattackBlk$\uparrow$} &
\textbf{CDR$\downarrow$} &
\textbf{TTR$\downarrow$} &
\textbf{ActProg$\uparrow$} \\
\midrule
Full (G+P+E) &
\textbf{0.95} &
\textbf{0.91} &
\textbf{1.00} &
\textbf{0.92} &
\textbf{1.00} &
\textbf{1.00} &
\textbf{0.00} &
18.57 &
0.34 \\
w/o E &
0.83 &
0.79 &
0.93 &
0.82 &
0.70 &
0.93 &
0.30 &
16.60 &
0.34 \\
w/o E+G &
0.69 &
0.63 &
0.53 &
0.84 &
\textbf{1.00} &
0.73 &
\textbf{0.00} &
10.80 &
0.38 \\
w/o E+G+P &
0.60 &
0.53 &
0.44 &
0.83 &
0.89 &
0.47 &
0.08 &
\textbf{7.40} &
\textbf{0.42} \\
\bottomrule
\end{tabular*}
\end{table*}

\subsection{RQ4: Design Attribution}
\label{subsec:rq4}

\noindent\textbf{Protocol and ablations.}
RQ4 evaluates the core components through cumulative ablations on STAIR-Bench.
G, P, and E denote GAS, the Stage-Specialized Agent Planner, and the Experience Database.
Because the database is indexed by GAS and the planner uses both GAS and retrieved experience, we compare four progressively simplified configurations: Full (G+P+E); w/o E, removing experience retrieval; w/o E+G, further replacing GAS with textual observations and recent action history; and w/o E+G+P, further disabling stage dispatch, stage working memory, and stage-based tool filtering, with the LLM planning directly from textual observations and recent action history.
All configurations use the main experiment's model, test set, tools, and 30-minute wall-clock limit; adjacent comparisons measure each component's incremental contribution in the remaining configuration.
The database is constructed from STAIR-Bench training ranges and provides effective and high-risk experience at test time, without incorporating test trajectories.

\noindent\textbf{Results.}
Table~\ref{tab:ablation} shows that progressively removing components reduces ODS from 0.95 to 0.83, 0.69, and 0.60, and Recovery from 0.91 to 0.53.
Attack suppression and replay blocking also consistently decline.
The later ablations shorten TTR to 10.80 and 7.40 and raise ActProg to 0.38 and 0.42 relative to Full's 18.57 and 0.34, yet overall response quality deteriorates: shorter trajectories and more local action progress do not imply complete recovery.

\noindent\textbf{Analysis.}
Compared with Full, w/o E reduces Recovery from 0.91 to 0.79 and SCS from 1.00 to 0.70, while increasing CDR from 0.00 to 0.30, indicating losses in recovery completeness and response safety.
Effect-validated historical records associate actions with their applicable states and actual consequences, supplying effective operations and risk warnings that help advance recovery while preserving services.
Further removing GAS reduces AttackSup from 0.93 to 0.53 and ReattackBlk from 0.93 to 0.73: even with stage planning, textual observations and recent history do not fully suppress attacks.
GAS continuously links affected assets, attack paths, and action feedback, keeping discovered but unresolved conditions available to subsequent decisions and supporting response across steps and hosts.
Further removing stage planning reduces Recovery from 0.63 to 0.53 and ReattackBlk from 0.73 to 0.47, showing the joint contribution of stage dispatch, stage memory, and scoped tools.
These mechanisms organize actions and their prerequisites around the current recovery objective, helping planning cover unfinished response requirements.

\noindent\textbf{Representative case.}
An SMTP propagation incident illustrates these differences.
The source, \texttt{infected-host}, sends an overlong \texttt{RCPT TO} request to \texttt{mail-relay:587} and retains a resident beacon; related logs and spool files reside on the mail host.
In the selected Full run, experience from similar incidents guides initial checks of the port, logs, propagation payload, and beacon; GAS maintains cross-host objects and their response states; and stage planning organizes investigation, preservation, eviction, and restoration objectives.
Without E, the system still maintains object states but explores investigation and recovery operations itself, repeatedly checking whether SMTP resumes listening.
Without E+G, mail-host logs and spool files are repeatedly classified as unconfirmed, skipping preservation, while some source-host objects enter cleanup before preservation is complete; this run has ReattackBlk of 0.00.
Further removing stage planning leads to deletion before attempted preservation and termination before recovery checks are complete.
These representative behaviors expose the roles of experience guidance, state continuity, and objective organization, providing process-level evidence for the aggregate ablation differences.

\subsection{Failure Analysis}
\label{subsec:failure-analysis}

STAIR can leave response requirements unmet even after successfully suppressing an attack.
In a representative Shellshock run, AttackSup and ReattackBlk both reach 1.00, but the final \texttt{hardening\_sudoers\_audited} check fails, indicating that the requirement to audit privilege configuration remains unmet.
This result shows that blocking the original attack does not establish that all hardening requirements have been satisfied.

This omission is consistent with STAIR's reliance on the model to identify response objectives.
GAS maintains incident state from alerts, tool observations, and execution feedback, keeping discovered but unresolved objectives available to subsequent decisions; it cannot ensure that the model identifies all relevant recovery requirements.
Stage planning organizes actions around response objectives such as hardening, but the specific audit scope and completion judgments still depend on the model, while historical experience provides guidance.
Thus, related configuration checks may still be overlooked even after known attack paths have been effectively addressed.
STAIR improves state continuity and objective organization during response, but complete remediation still depends on the model identifying sufficient response requirements from visible evidence.

\section{Discussion}

\noindent\textbf{Scope and Integration with Existing Workflows.}
STAIR targets long-horizon response across assets and stages, tracking unresolved objectives and revising actions as evidence arrives.
For stable, low-risk one-step tasks, static playbooks or rule-based workflows may remain simpler and more cost-effective~\cite{cacao_playbooks,bridges2023testing,stevens2022ready}.
It can serve as a planning layer integrated with existing tools and procedures, selecting actions from the incident state while established tools execute them.
Transferable experience can guide investigation, containment, and recovery checks~\cite{woods2023lessons,zhao2024expel}.

\noindent\textbf{Organizational Adaptation and Execution Control.}
Enterprise deployment requires configuring asset and identity mappings, telemetry formats, tool interfaces, recovery objectives, and service-continuity constraints~\cite{kokulu2019matched,jones2023work,akbari2024requirements}.
Domain experts define these requirements and operational boundaries; STAIR handles incidents within that scope.
Cross-environment state alignment and experience adaptation remain research directions for reducing configuration effort and enabling reuse.

Response actions may involve host isolation, access-control changes, file cleanup, or service restart~\cite{scarfone2008sp, zhang2025survey}.
STAIR constrains operations through structured actions, stage-scoped tools, and the Execution Harness, using historical high-risk actions and their side effects as planning warnings.
Organization-specific approval policies can allow routine investigation to proceed automatically and require human review for operations affecting critical services~\cite{shaked2023operations,singh2025llms}.
Future work can align these policies with asset criticality, action impact, and incident state within existing SOC roles.

\noindent\textbf{Data Privacy and Deployment Choices.}
Our experiments use cyber-range data; enterprise deployment requires handling asset identifiers, logs, and response records under organizational data-management requirements.
For external model calls, sanitization at the model-call boundary can use local mappings to preserve entity correspondence across alerts, GAS, and retrieved experience.
Where data cannot leave the organization, an internal model endpoint can be used.
Preserving the associations needed for state reasoning and experience transfer during sanitization remains a research direction.

\section{Conclusion}

This paper presents STAIR, an end-to-end agentic planning framework for long-horizon incident response in live software systems.
Through a shared, evolving GAS, STAIR connects stage assessment, action selection, effect-validated experience, and execution feedback to track unresolved response objectives.
STAIR achieves more complete recovery and stronger attack suppression than representative baselines on STAIR-Bench and outperforms the Claude Code baseline with the same model backend in CAGE4 under ongoing attacks and benign activity.

\section{Data Availability}

The STAIR implementation, STAIR-Bench cases, and evaluation scripts are available in an anonymous replication package at \url{https://anonymous.4open.science/r/stair}.

\bibliographystyle{ACM-Reference-Format}
\bibliography{reference}

\end{document}